\documentclass[]{IEEEtran}
\usepackage{algorithm}
\usepackage{algpseudocode}
\usepackage{lipsum}
\usepackage{amsmath}
\usepackage{amsfonts}
\usepackage{amssymb}
\usepackage{graphicx}
\usepackage{cite}
\usepackage{amsthm}
\usepackage{epstopdf,epsfig}
\usepackage{xcolor}
\usepackage{balance}
\usepackage{colortbl}
\usepackage{url}
\usepackage{algpseudocode}
\usepackage{booktabs}
\usepackage{algorithmicx}
\usepackage{pbox}
\usepackage{subcaption}
\usepackage{dblfloatfix}
\usepackage{array}

\newtheorem{theorem}{Theorem}
\newtheorem{definition}{Definition}

\begin{document}
\newcommand\sForAll[2]{ \ForAll{#1}#2\EndFor}
\newcommand\sIf[2]{ \If{#1}#2\EndIf} 

\title{A Game-Theoretic Analysis of Shard-Based Permissionless Blockchains}
\author{\IEEEauthorblockN{Mohammad Hossein Manshaei\textsuperscript{$\star$}, Murtuza Jadliwala\textsuperscript{$\dagger$}, Anindya Maiti\textsuperscript{$\ddagger$}, \textit{and} Mahdi Fooladgar\textsuperscript{$\star$}}\\
\IEEEauthorblockA{\textsuperscript{$\star$}Department of Electrical and Computer Engineering, Isfahan University of Technology, Iran\\
\textsuperscript{$\dagger$}Department of Computer Science, University of Texas at San Antonio, USA\\
\textsuperscript{$\ddagger$}Institute for Cyber Security, University of Texas at San Antonio, USA\\
Email: manshaei@cc.iut.ac.ir, murtuza.jadliwala@utsa.edu, a.maiti@ieee.org, m.fooladgar@ec.iut.ac.ir}}

\maketitle

\begin{abstract}
Low transaction throughput and poor scalability are significant issues in public blockchain consensus protocols such as Bitcoins. Recent research efforts in this direction have proposed shard-based consensus protocols where the key idea is to split the transactions among multiple committees (or shards), which then process these shards or set of transactions in parallel. Such a parallel processing of disjoint sets of transactions or shards by multiple committees significantly improves the overall scalability and transaction throughout of the system. However, one significant research gap is a lack of understanding of the strategic behavior of rational processors within committees in such shard-based consensus protocols. Such an understanding is critical for designing appropriate incentives that will \emph{foster cooperation} within committees and \emph{prevent free-riding}. In this paper, we address this research gap by analyzing the behavior of processors using a game-theoretic model, where each processor aims at maximizing its reward at a minimum cost of participating in the protocol. We first analyze the Nash equilibria in an $N$-player static game model of the sharding protocol. We show that depending on the reward sharing approach employed, processors can potentially increase their payoff by unilaterally behaving in a \emph{defective} fashion, thus resulting in a social dilemma. In order to overcome this social dilemma, we propose a novel \emph{incentive-compatible reward sharing} mechanism to promote cooperation among processors. Our numerical results show that achieving a majority of \emph{cooperating} processors (required to ensure a healthy state of the blockchain network) is easier to achieve with the proposed incentive-compatible reward sharing mechanism than with other reward sharing mechanisms.
\end{abstract}

\section{Introduction}
\label{sec:Intro}

A \emph{blockchain} is an append-only, immutable distributed database that records a time-sequenced history of facts called \emph{transactions}. Transactions are typically grouped into \emph{blocks}, and the blockchain protocol enables the construction and maintenance of consistent copies of the cryptographic \emph{hash-chain} of blocks in a distributed fashion. The first blockchain protocol was introduced in 2009 by Satoshi Nakamoto to support the \emph{Bitcoin} cryptocurrency \cite{bitcoin:2009}. A key aspect of this protocol is the \emph{consensus} algorithm (also sometimes referred in the literature as \emph{Nakamoto consensus}) which enables agreement among a network of \emph{processors} or \emph{miners} on the state of the blockchain (identified by its cryptographic digest), under the assumption that a fraction of them could be malicious or faulty. In addition to this, as Bitcoin's blockchain is \emph{permissionless}, i.e., no trusted infrastructure to establish verifiable identities for processors exists or is assumed, consensus on the blockchain's state cannot be achieved using standard distributed Byzantine fault-tolerant consensus algorithms in the literature. In such a permissionless setting, the blockchain protocol selects (randomly and in an unbiased fashion) one processor once every 10 minutes on average (\emph{epoch}), and this selected processor gets the right to commit (or append) a new block onto the blockchain. The network (other processors) \emph{implicitly} accept this block by building on top of it in the next epoch or reject it by building on top of some other block in the hash-chain.

Consensus in Bitcoin is thus \emph{long-term}, i.e., a block is said to be included in the blockchain if it is part of the longest valid blockchain and has received a significant number of confirmations\footnote{Number of blocks added on top of the block in question in the longest valid blockchain.}. The Bitcoin protocol uses a \emph{Proof-of-Work (PoW)} mechanism to select the leader (processor with the right to commit a block) in each epoch in an unbiased fashion, which is nothing but a \emph{hash puzzle} that each processor attempts to solve - one that succeeds is selected and gets the right to propose the next block. As PoW involves significant computation, Bitcoin's protocol includes a \emph{reward mechanism} to incentivize processors to compete (in a fair fashion) and to behave honestly. As of July 2018 \cite{allcrypto}, there were a total of 1624 cryptocurrencies, a significant number of which use the same code base as Bitcoin or are directly inspired by Bitcoin's distributed consensus algorithm. The use of blockchains and blockchain-based distributed consensus, however, is not just restricted to cryptocurrencies. Systems that can host and execute arbitrary distributed applications (commonly referred to as ``\emph{smart contracts}") over a single public permissionless hash-chain, for example, \emph{Ethereum} \cite{ethereum}, have also become popular. Such systems also employ a Bitcoin-like Proof-of-Work based consensus algorithm and a related cryptocurrency (e.g., Ether in Ethereum) to incentivize processors or miners to participate honestly in the consensus process.

Despite its tremendous popularity, one significant shortcoming of Bitcoin's consensus protocol (and of similar public permissionless blockchain systems) is its low transaction throughput and poor scalability. With an average inter-block time of 10 minutes and a maximum block size of 10 MB, Bitcoin's transaction rate is currently only 7 transactions per second \cite{bitcoinscalability}. Similarly, Ethereum can support only roughly 20 transactions per second. This is significantly lower than the transaction rates afforded by centralized transaction processing systems. For instance, PayPal can process more than 450 transactions per second while VisaNet can process anywhere between 1667 and 56,000 transactions per second \cite{bitcoinscalability}. It is clear that the current Bitcoin and Ethereum transaction rates are not sufficient for many practical applications, and thus, there have been significant efforts towards improving their transaction throughputs, for example, BIP \cite{bip102} and Bitcoin-NG \cite{eyal2016bitcoin} for Bitcoin and Raiden \cite{raidennetwork} for Ethereum.

Similarly, there have been other significant efforts within the research community towards improving the transaction throughput and scalability of public permissionless blockchain protocols in general. One key outcome of this line of research is \emph{sharding} \cite{luu2016secure,zamanirapidchain,kokoris2017omniledger}, which proposes to periodically partition the network of processors (in an unbiased fashion) into smaller \emph{committees}, each of which processes a disjoint set of transactions (also called a \emph{shard}\footnote{Note that the committees are working inside shards in these protocols. Hence, we use the two terms interchangeably in the paper.}) in parallel with other committees. As each committee is reasonably small, it can run a classical Byzantine consensus protocol such as PBFT \cite{Castro:2002} to agree on a set of transactions rather than the traditional Nakamoto consensus of Bitcoin, thus increasing the overall transaction throughput of the system. Although the idea of parallelizing the tasks of transaction processing and reaching consensus (on a set of transactions) by partitioning the processor network into committees is promising, existing sharding proposals \cite{luu2016secure,zamanirapidchain,kokoris2017omniledger} fail to clarify how processors will be incentivized to honestly participate and discharge their committee duties. 

Two facts about existing sharding protocols are relevant to this discussion and should be highlighted: (i) the intra-committee consensus algorithms (e.g., PBFT) employed by existing protocols are inherently \emph{fault-tolerant}, i.e., they will operate correctly even in the presence of a certain number of faulty or non-participating committee members, and (ii) the agreed (or consensus) set of transactions within each committee is required to be ratified (or signed) by only a majority of the committee members in order for those to be included into a block.
Now as participation in committee tasks (such as transaction validation, signature creation, etc.) impose a \emph{cost} on processors, it is possible that \emph{rational} processors may choose not to participate in these tasks (and get away with it as the protocol may still succeed at the end) if their remuneration is not appropriately determined. For example, if each processor within a committee is equally remunerated, a rational processor may choose to \emph{free-ride}, i.e., get paid without participating in any committee work.  
\emph{In summary, one key research gap in this line of research is a lack of understanding of the strategic behavior of rational processors in shard-based consensus protocols for public permissionless blockchains.} Such an understanding is critical for designing appropriate incentives that will \emph{foster cooperation} within committees and \emph{prevent free-riding}. Our goal in this paper is to address this research gap.

In line with the above goal, we first model shard-based protocols, and the interaction between processors in such protocols, using a static non-cooperative game by systematically quantifying processor strategies in such a game and the resulting payoffs. We show that in such a setting, if the total reward (received at the end of the game when a new block is successfully committed to the blockchain) is equally or uniformly distributed among all the participating processors, then the resulting strategic interactions can be characterized using a game with social dilemma, such as a \emph{public goods game}. Consequently, we show that not participating in the committee tasks (by all processors) is a Nash equilibrium of the game. We further show that it is impossible to enforce a cooperative Nash equilibria in this setting unless certain improbable conditions are met. Hence, we extend the current game model by considering \emph{fair sharing} of rewards, instead of equal sharing, where processors receive benefits only if they have cooperated within their shards. In this new system, we derive the Nash equilibria and conditions under which such an equilibria can be achieved. Although this game is still a public goods game, we were able to establish conditions for achieving cooperation by processors towards executing the committee tasks in this game. These conditions can be derived and verified by processors before they decide on their strategy to \emph{cooperate} or \emph{defect} in the game. Our results show that it is possible to achieve a cooperative equilibium in such a fair reward sharing system. Finally, we design the \emph{incentive-compatible} reward sharing protocol that further improves upon the fair sharing protocol by introducing a shard coordinator who can guide individual processors to follow the optimal strategy (cooperate or defect), based on a preview of the shard's consensus status in each epoch. Our numerical analysis show that the \emph{incentive-compatible} protocol can outperform both the uniform and fair reward sharing protocols. To the best of our knowledge, this paper is the first to investigate the selfish behavior of processors, and its effect, in shard-based permissionless blockchains.

The rest of the paper is organized as follows. In Section~\ref{sec:sysmodel}, we discuss the state of the art and present a generic system model for shard-based blockchain protocols, considering rational processors. In Section~\ref{sec:results}, we present the game model and investigate all possible Nash equilibria under different reward sharing schemes. In Section~\ref{sec:MD}, we describe the proposed \emph{incentive-compatible} reward sharing protocol, followed by numerical evaluations presented in Section~\ref{sec:Sim}. Related research efforts efforts have been outlined in Section~\ref{sec:related}. We conclude the paper in Section~\ref{sec:conclusions}.

\section{System Model}
\label{sec:sysmodel}
In this section, we first generically outline details of a shard-based approach for achieving consensus in permissionless blockchains. Then, we formally outline the various costs involved for participating processors. Lastly, we clarify the rationality assumptions related to the processors.

\subsection{Shard-based Consensus Protocol}
Consider a network of $N$ processors participating in a public permissionless blockchain. Processors in such a network do not have an identity assigned by a trusted third-party or a public-key infrastructure, i.e., they use self-generated pseudonyms as transient identifiers. For simplifying the exposition, we assume that all processors are \emph{similar} to each other in terms of \emph{computational capabilities}. Further, we assume that all processors are \emph{honest}, but \emph{selfish} (more details on this will follow in section \ref{sec:rationality}).

Let time be divided into fixed-sized \emph{epochs}. The network accepts \emph{transactions} in \emph{blocks}, i.e., at the end of each epoch the network accepts and commits a new block of transactions. Any block $B$ is composed of (or can be partitioned into) $k$ disjoint sets of transactions $B_i$, where $B_i$ can be empty. Each such \emph{disjoint set} $B_i$ is referred to as a \emph{shard} and can be defined based on some property(ies) of transactions within that set, for example, least significant bits of the transaction hash.  The number of shards ($k$) is a variable quantity and can grow linearly with the size of the network. The network determines a binary \emph{validation function} $V$, which takes as an input a transaction (belonging to any shard) and any other data representing the current state of the blockchain and outputs whether the input transaction is valid or not, and all processors have access to such a function $V$. 

\begin{figure}[]
	\centering
	\includegraphics[width=\linewidth]{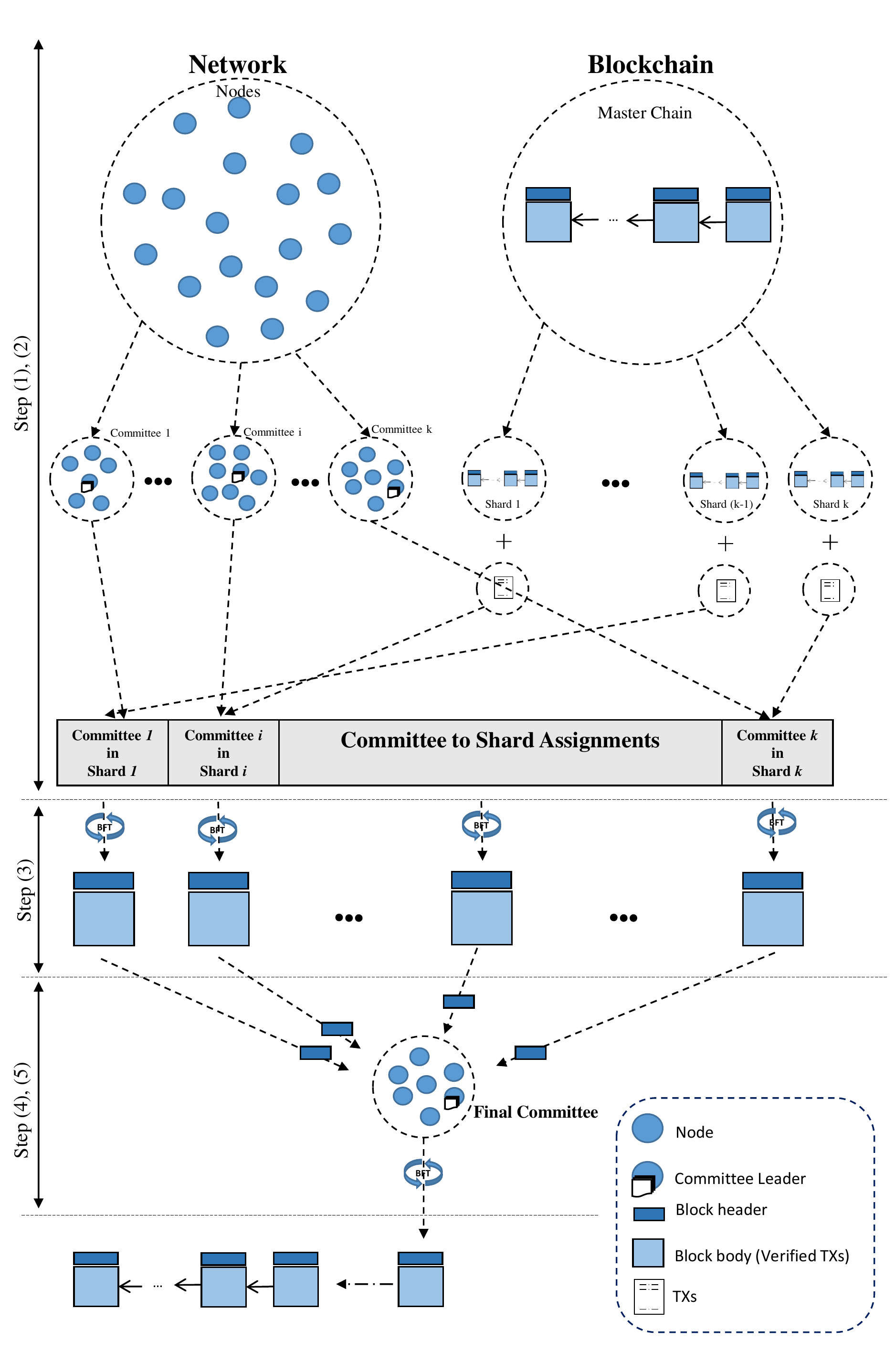}
	\caption{Conceptual view of a shard-based consensus protocol.}
	\label{fig_sysmodel}
\end{figure}

Now, given the above, a \emph{sharding} or \emph{shard-based protocol} is a protocol which is run among the processors and which outputs (at the end of each epoch) a block $B$ containing $k$ disjoint shards $B_i$ such that all honest processors agree on $B$ with a very high probability and all transactions within the block $B$ are valid (i.e., all transactions satisfy the validation function $V$). The protocol does this by splitting the network of processors into multiple disjoint \emph{committees}, where each committee processes (validates and agrees on) a separate shard ($B_i$). The main steps in the protocol execution during each epoch is illustrated in Figure \ref{fig_sysmodel}. Below, we summarize the main steps involved in sharding by outlining a classical protocol called \emph{Elastico} \cite{luu2016secure}. Recent research efforts such as Omniledger \cite{kokoris2017omniledger} provide some enhancements and additional functionalities to the original sharding proposal in Elastico, but the key idea of partitioning the transactions into disjoint shards and assigning a committee of processors to process each shard in parallel remains the same in all shard-based protocols.

A sharding protocol proceeds in epochs and in each epoch the processors execute the following steps (in this order) \cite{luu2016secure}:
\begin{enumerate}
\item \emph{Committee Formation}: First, each processor attempts to generate a publicly verifiable identity by solving some \emph{Proof-of-Work (PoW) puzzle}. In other words, each processor uses the solution of a PoW hash puzzle (i.e., the message digest that lies within the pre-determined target) as an identity in that epoch. There are two advantages of using a PoW puzzle for identity creation: (i) network (other processors) can verify the identity and, (ii) number of malicious sybils can be limited due to the computation involved in solving the puzzle. Each processor is then assigned to a committee corresponding to its established identity (say, using the $s$ least significant bits of the identity). Moreover, each committee processes a distinct shard based on this $s$-bit identifier.
\item \emph{Overlay Setup}: Next is the \emph{community discovery} step where processors discover identities of other processors in their committee by communicating with each other. The outcome of this step is a \emph{fully-connected overlay} for each committee in the network.
\item \emph{Intra-Committee Consensus}: Next, processors run a standard \emph{byzantine agreement protocol} such as \emph{PBFT} \cite{castro1999practical} within their committees to agree on a set of transactions. Each committee then sends its consensus set of transactions $B_i$ (or shard) to a \emph{final committee} for inclusion in the new block $B$ at the end of current epoch. In order to be considered by the final committee, each shard $B_i$ needs to be signed by a \emph{simple majority}, i.e., by at least  $\frac{c}{2} + 1$ processors for a committee of size $c$.
\item \emph{Final Consensus}: A final committee (chosen based on a designated $s$-bit final committee identifier) then takes the consensus shards ($B_i$) from the previous step and merges these to create a final block $B$, creates a cryptographic digest or hash of $B$ and broadcasts it to the rest of the network. 
During the merge operation, each processor in the final committee first validates that each shard $B_i$ is signed by at least $\frac{c}{2} + 1$ processors in the correct committee and then computes a union of all the shards to form the block $B$. After each processor in the final committee computes a union in this fashion, they then collectively run a byzantine agreement protocol such as PBFT\cite{castro1999practical} to arrive at a consensus on the final block $B$. The cryptographic digest of the final consensus block $B$ needs to be signed by a simple majority of the final committee before it can be broadcast on the network.
\item \emph{Randomness Generation for Next Epoch}: In the final step of the protocol, the final committee generates a set of \emph{random strings} and broadcasts it to the network. These random strings are used by the processors in the identity creation and committee formation tasks of the \emph{next epoch}.
\end{enumerate}

\subsection{Processor Costs}
\label{sub:Costs}
We now characterize the costs (including, computation and communication costs) borne by the processors in each time epoch due to their participation in the sharding protocol. It should be noted that our goal here is not to arrive at a precise quantification of these costs, rather to characterize them such that they could be used to analyze the strategic behavior of processors while participating in the protocol.
The protocol steps in each epoch, as outlined in the previous section, can be basically grouped into two phases: (1) \emph{organization phase} and, (2) \emph{committee participation phase}. During the organization phase, the processors create identities using PoW puzzles, form committees and identify other processors in their committee (i.e., execute steps 1 and 2 in the protocol above), whereas in the committee participation phase the processors validate their respective shards and arrive at an agreement with other committee members (i.e., execute steps 3, 4 and 5 in the protocol above). 

It should also be clear from the protocol description above that the organization phase precludes the committee participation phase, and it is \emph{required} or \emph{mandatory}. In other words, if a processor does not go through the organization phase, it does not have an identity nor it gets assigned to a committee, and so it cannot take part in the committee participation phase. Similarly, it should also be clear that the committee participation phase is not mandatory for processors, i.e., a processor could choose to create a verifiable identity and be assigned to a committee, but may choose not to participate in tasks such as shard validation and intra-committee consensus. If some processors do not take part in the committee participation phase, it does not mean that the protocol will fail. The inherent fault-tolerance of intra-committee consensus protocols such as PBFT and the simple majority rule employed in intra-committee voting implies that a certain number of non-participation can be tolerated by the protocol. For the sake of convenience, we assume that if more than half ($> \frac{c}{2}$) of the processors within a committee of size $c$ do not participate in the committee participation phase, the entire protocol for that epoch fails, i.e., no new block is proposed in that epoch. 

Thus, we can characterize the total cost for a processor to participate in an epoch of the sharding protocol based on the cost for executing the above two phases. For the organization phase, let's assume that a processor bears a cost $c^m$, which we refer to as the \emph{mandatory cost}. It should be noted that $c^m$ is a fixed cost and is independent of the number of transactions processed by the processor. Moreover, as solving the PoW puzzle is the most significant activity during the organization phase, $c^m$ can be approximated using the current difficulty of the PoW puzzle and the average computational power of all the processors. 

Accordingly, for executing the committee participation phase let's assume that a processor bears an \emph{optional cost} $c^o$, depending on whether the processor fully participates in it or not. Unlike the mandatory cost, the optional cost $c^o$ has two components: (i) a \emph{fixed component} and, (ii) a \emph{transaction-dependent component}. During the committee participation phase, a processor performs activities such as participation in intra-committee consensus the cost of which can be bounded by a fixed average cost \cite{castro1999practical}. We represent all these per-processor fixed costs during the committee participation phase as $c^f$. Another activity during this phase that all processors are expected to perform is verifying the validity of all outstanding transactions (they have received) within their respective shards by using the validation function $V$. Depending on the complexity of the validation function $V$, this can be a significant cost (to a processor) which also depends on the number of outstanding transactions being validated. We represent the cost to validate each transaction using $V$ by $c^v$. Hence, we can compute the total optional cost $c_i^o$ for a processor $P_i$ as:
\begin{equation}
c_i^o= c^f + |x_i^j|c^v
\end{equation}
where $x_i^j$ is the vector of transactions received and validated by processor $P_i$. The average per-processor cost ($c_i^t$) for participation in each epoch of the shard-based protocol can thus be characterized as $c_i^t=c^m+c^f+|x_i^j|c^v$. 

One point that needs further clarification is why a processor may choose not to execute the committee participation phase after executing the organization phase. Our rationality assumption, which we describe next, provides this clarification.

\subsection{Rationality Assumption}
\label{sec:rationality}
Earlier research efforts on sharding \cite{luu2016secure,kokoris2017omniledger} have assumed a \emph{byzantine adversary} where processors controlled by the adversary can be \emph{arbitrarily malicious}, i.e., malicious processors could arbitrarily deviate from the correct execution of the protocol or could arbitrarily drop protocol messages. In this work, however, we assume that processors are \emph{honest} but \emph{selfish}. In other words, processors do not arbitrarily deviate from protocol execution or drop protocol messages, but decide against participation in the protocol only when there is an incentive (financial or otherwise) to do so. Let us further provide a brief intuition of the notion of rationality in this setup. All processors receive some rewards if the protocol execution in an epoch is successful, for example, in terms of block rewards, transaction fees, etc. The precise nature of rewards depend on the specific system or application that the blockchain protocol enables. Moreover, as discussed in the earlier section, all processors bear some costs for fully participating in both phases of the protocol. The total \emph{benefit} or \emph{payoff} received by processors in each epoch is the difference between the obtained reward and the spent costs in that epoch. A selfish (or rational) processor will always choose a protocol participation strategy that improves its benefit or payoff. If a processor does not execute the organization phase, it does not get any reward as it is not a part of any committee. However, a rational processor's strategy could be to execute the organization phase but refrain from the committee participation phase. Such a selfish strategy saves on the optional cost $c^o$ and may result in a reward if enough other processors participate fully, and thus may provide more benefit or payoff to the rational processor. We assume that a rational processor will always choose such a selfish strategy which provides more benefit or payoff, if it exists. In summary, the goal of each processor is to maximize its individual payoff (received at the end of each epoch), without maliciously trying to deviate or disrupt the protocol. We assume that processors do not collude/coordinate in order to jointly maximize their combined utility.
\section{Shard-Based Blockchain Game}
\label{sec:results}

\begin{table}[t]
\caption{List of Symbols.}
\begin{center}
\begin{tabular}{c | l } \hline
{ \textbf{Symbol}}  & { \textbf{Definition}}   \\  \rowcolor[gray]{.9} \hline
$k$		& Number of shards (or committees) \\
$N$		& Number of processors \\ \rowcolor[gray]{.9}
$x_i^j$	& Vector of received transactions by processor $i$ in shard $j$ \\
$y^j$		& Vector of transactions submitted by shard $j$ to Blockchain \\\rowcolor[gray]{.9}
$c$		& Minimum number of processors in each committee \\ 
$\tau$	& Required number of processors in shard for consensus \\\rowcolor[gray]{.9}
$r$		& The benefit for each transaction \\ 
 $b_i$		& Benefit of processor $i$ after adding the block\\ \rowcolor[gray]{.9}
$c^t_i$	& Total cost of computation for processor $i$\\
$c^o$		& Total optional costs in each epoch \\ \rowcolor[gray]{.9}
$c^m$	& Mandatory costs in each epoch to enter the shard \\ 
$c^v$		& Cost of transaction verification\\\rowcolor[gray]{.9}
$c^f$		& Fixed costs in optional cost\\
$BR$		& Block Reward\\\rowcolor[gray]{.9}
$l_j$		& Number of cooperative processors in each shard\\
$L$		& Total number of cooperative processors in all shards\\\rowcolor[gray]{.9}
$C_j^{l_j}$	& The set of all cooperative processors in shard $j$\\
$D_j^{n-l_j}$	& The set of all defective processors in shard $j$ \\\rowcolor[gray]{.9}
$C^L$	& The set of all cooperative processors \\
$D^{N-L}$	& The set of all defective processors \\ \hline
\end{tabular}
\end{center}
\label{tab:Symbols}
\end{table}

In this section, we present the game-theoretic aspects of a shard-based blockchain protocol with multiple processors in a honest but selfish environment. We first introduce a non-cooperative $N$-Player game model that we refer to as the \emph{shard-based blockchain game} $\mathbb{G}$. Upon starting an epoch $t$, processors must decide whether to collaborate with each other, verify transactions, and make a block to be appended to the chain (i.e., take part in the community participation phase), after the organization phase as we addressed in the previous section. The key point of the game-theoretic analysis is to consider the computation costs for processors who verify transactions and participate in consensus mechanism, as presented in Section \ref{sub:Costs} and \ref{sec:rationality}, and the total benefits when they agree on a valid block. Therefore, using a game-theoretic analysis, we investigate whether block generation can emerge in such a non-cooperative system. By means of our game model and the related analysis, we would like to show that with a uniform distribution of rewards in these protocols,  the interactions between processors fall in a category of games, where there exists a social dilemma of all-defection behavior. We then propose a novel reward sharing protocol and address the conditions for having a new class of equilibrium, where a subset of processors will be forced to cooperate. Table \ref{tab:Symbols} summarizes the notation used throughout the paper.

\subsection{Game Model}
Game theory allows for modeling situations of conflict and for predicting the behavior of participants when they interact with each other. In our \emph{shard-based blockchain game} $\mathbb{G}$, processors must decide upon joining the shards whether to cooperate and contribute to optional costs (as addressed in Section \ref{sub:Costs}) or not. We model the \emph{shard-based blockchain game} as a static game, because all processors must choose their strategy simultaneously, after they have joined the shards.  This modeling decision also keeps our analysis tractable, while conforming to a simple model of processor rationality. The game $\mathbb{G}$ is defined as a triplet $(\mathcal{P}, \mathcal{S}, \mathcal{U})$, where $\mathcal{P}$ is the set of players, $\mathcal{S}$ is the set of strategies and $\mathcal{U}$ is the set of payoff values. We also assume that at any time epoch $t$, a game is played among all the processors in all shards, because the benefits of successfully adding a block is shared among all processors.

{\bf $\bullet$ Players ($\mathcal{P}$):} The set of players $\mathcal{P} = \{P_i\}_{i=1}^{N}$  corresponds to the set of processors who have already joined shards in a given epoch time $t$. In fact, all $N$ processors must have already performed PoW and paid the mandatory costs $c^m$.
Considering the number of shards in our system model, i.e., $k$, we conclude that each shard has $n=N/k$ committee members. During this epoch time in our game, we assume that each processor $P_i$ in shard $j$ receives the vector  $x_i^j$ of transactions to verify and participate in the consensus algorithm. 

\begin{figure}[t]
	\centering
	\includegraphics[scale=0.5]{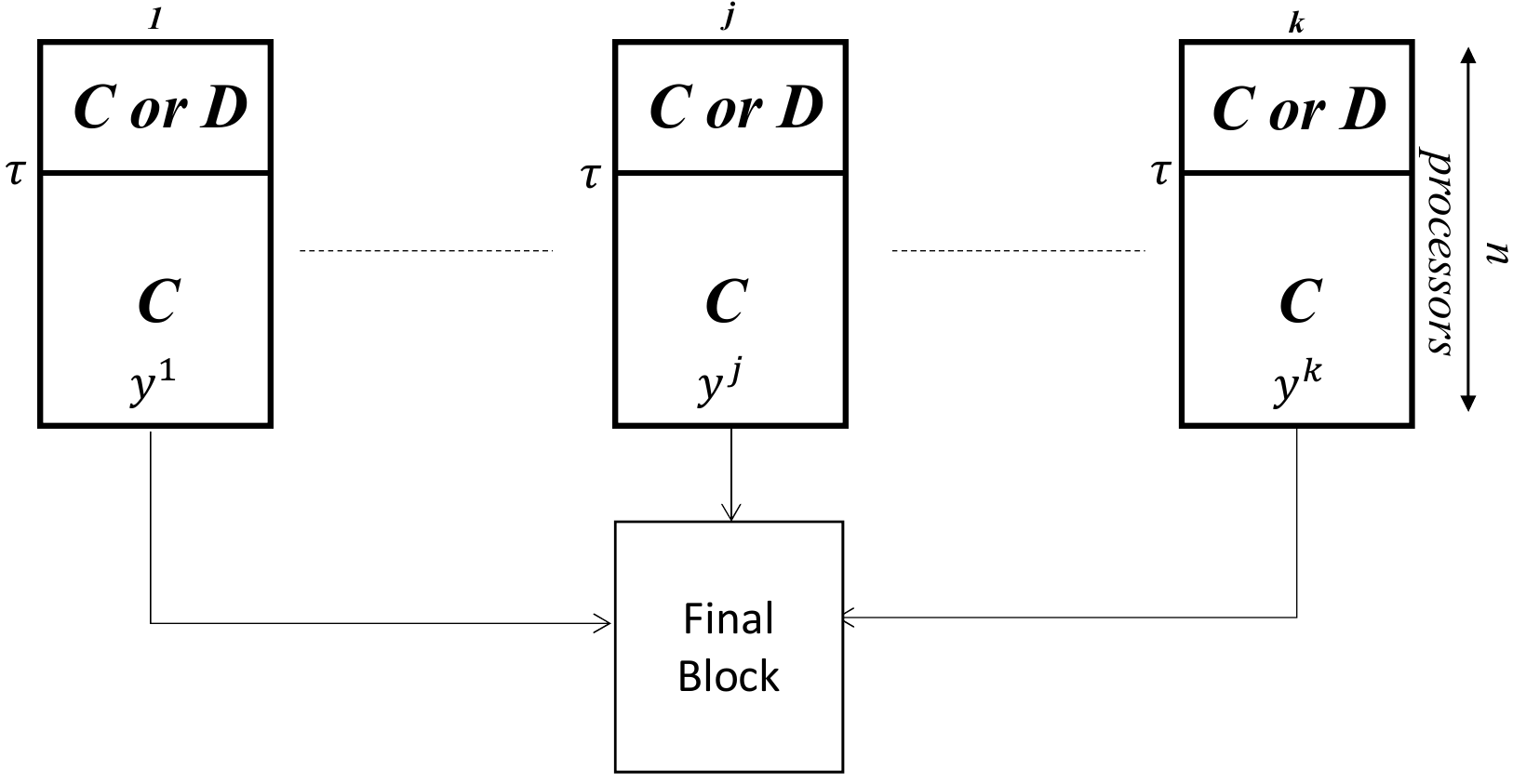}
	\caption{In each shard at least $\tau$ processors among $n$ processors must be cooperative to perform consensus algorithm. Each Shard $j$ submits the final $y^j$ vector of transactions to make the final block.}
	\label{fig_GameModel}
\end{figure}

As it is shown in Figure \ref{fig_GameModel}, we also assume that to perform a consensus algorithm in each shard we need at least $\tau$ processors who agree on a given list of transactions. For example, in Elastico protocol which uses PBFT, $\tau$ is equal to $\frac{2}{3} n$. Finally, $y^j, \; j\in \{1, ..., k\}$ represents the result of the consensus algorithm including the list of transactions that would be added to the blockchain by shard $j$. 

{\bf $\bullet$ Strategy ($\mathcal{S}$):} Each processor $P_i$ can choose between two moves $s_i$: \emph(i) \emph{Cooperate} $C$, or \emph{(ii)} \emph{Defect} $D$. Hence the set of strategies in this game is $\mathcal{S} =\{C, D\}$. The strategy of processor $P_i$ determines whether $P_i$ participates in all optional tasks presented in Section \ref{sec:sysmodel} or not. In particular, if processor $P_i$ plays $C$, it will accept and verify all received transactions. In this case, it also cooperates in all consensus algorithms and incurs cost $c^o$ for its participation. Contrary to a cooperative behavior, a given processor can refuse all transaction verifications and simply do nothing during the community participation phase (i.e., play $D$).

{\bf $\bullet$ Payoff ($\mathcal{U}$):} Without loss of generality, we assume that each transaction will make $r$ benefits, for example in form of a transaction or some other fee (a similar model exists in Bitcoin and other popular cryptocurrencies). Hence, the total benefit that a given shard $j$ can make from transactions is $b_j=r|y^j|,$ where $y^j$ is the set of verified, signed, and accepted transactions by shard $j$. This benefit term $r$ for the network is some function of the average \emph{transaction fee} included in the transactions and number of committee (shard) members that have processed the transaction. A precise quantification of $r$ is not trivial and is considered out of scope of this work. Finally, the total benefits that are made by all transaction fees in the final appended block can be calculated as $TF= r \sum_{j=1}^{k}|y^j|$.

Recall that the total cost of cooperation for processor $P_i$ is equal to $c_i^t=c^m+c_i^o=c^m+c^f+|x_i^j|c^v$ if the processor acts honestly and follows the protocol. All processors should pay the mandatory costs, i.e., $c^m$ in order to be in a committee and finally receive the reward. But they can avoid paying optional cost $c^o$, including the cost of verifications. In summary, we can divide processors into two groups of cooperative and defective processors, based on whether they contribute to optional tasks (i.e., play $C$ and pay $c^t$) or not (i.e., play $D$ and pay only $c^m$). Let $\mathcal{C}_j^{l_j}$ and $\mathcal{D}_j^{n-l_j}$ denote the sets of $l_j$ cooperating players and $n-l_j$ defecting players in a given shard $j$ with $n$ processors. Recall that in order to obtain a consensus transaction vector $y^j$ in a given shard $j$, $|\mathcal{C}_j^{l_j}|$ must be greater than or equal to $\tau$ ($|\mathcal{C}_j^{l_j}| \ge \tau$). 

After executing the protocol and inserting the computed block to the blockchain at the end of the epoch, we assume that the system receives two rewards. This assumption is motivated from the observation in current public blockchain applications such as Bitcoin and Ethereum. The first reward is a fixed reward for adding a new block, called the \emph{block reward} ($BR$). The current block reward for Bitcoin, for example, is 12.5 BTC \cite{bcexplorer}. The second reward is the sum of transaction fees which is equal to $TF= r \sum_{j=1}^{k}|y^j|$. Note that all following analyses are based on the assumption that each shard has already provided a non-empty $y^j$. Due to lack of clarity in shard-based blockchain proposals \cite{luu2016secure,zamanirapidchain,kokoris2017omniledger}, we assume that if one or more shards fail to provide a $y^j$ in an epoch, the system cannot compute and append a new block in that epoch.

If we assume that all processors receive an equal share of profits after block computation (i.e., the existing protocol), we can calculate the reward share for each processor as $$\frac{BR+r\sum_{i=1}^{k}|y^j|}{N}.$$ In other words, all processors receive equal share of the rewards from block reward and total transaction fees. Hence, we can compute the payoff of processor $P_i$ in shard $j$ by 

\begin{equation}\label{eq:PayoffC}
u_i^j(C) = b_i - c_i^t =\frac{BR+r\sum_{j=1}^{k}|y^j|}{N} - ( c^m+c^f+|x_i^j|c^v), 
\end{equation}

if we assume that the processor $P_i$ was cooperative, i.e. $P_i \in \mathcal{C}_j^{l_j}$. Similarly, if $P_i$ is defective, i.e., $P_i \in \mathcal{D}_j^{n-l_j}$, its payoff would be:

\begin{equation}\label{eq:PayoffD}
u_i^j(D) =\frac{BR+r\sum_{j=1}^{k}|y^j|}{N} - c^m. 
\end{equation}

Considering the above calculated payoffs, we analyze the game $\mathbb{G}$ next.
  
\subsection{Game Analysis}

In order to get an insight into the strategic behavior of the processors, we apply the most fundamental game-theoretic concept, named Nash equilibrium, introduced by John Nash \cite{nash1951non}: 

\begin{definition}
In a Nash equilibrium strategy profile, none of the players can unilaterally change his strategy to increase his utility. 
\end{definition}

In other words, if in a non-cooperative game all strategies are mutual best responses to each other, then no player has any motivation to deviate unilaterally from the given strategy profile. Nash also proved that any finite game has at least one Nash equilibrium strategy profile. In non-cooperative game theory, \emph{Prisoner's dilemma} or PD, discovered by Flood and Dresher in 1950 and later formalized by Tucker \cite{Tucker1950}, is a classical 2-player game which shows why two rational individuals might not cooperate, even if it appears that the cooperative strategy is more beneficial for both of them (i.e., Pareto Optimality). In PD, each individual has two strategies of \emph{cooperation} and \emph{defection}, and the defection strategy strictly dominates the cooperation strategy. Hence, the only Nash equilibrium in PD, is a mutual defection. 

More than 20 years later, Hamburger defined the analogous $N$-player version of PD game in \cite{hamburger1973n}. This extension is called \emph{public good game} (PGG). In a PGG setting, each individual can cooperate and pay a contribution of $\alpha$ or defect and do not pay anything. Then all contributions would be summed and multiplied by a reward factor $\gamma >1$. Finally, the total reward would be distributed among all users equally, whether they have cooperated or defected. In other words, if $n$ agents out of $N$ cooperate, their payoff would be $\frac{\gamma \alpha n}{N}-\alpha$ and the defectors' payoff is $\frac{\gamma \alpha n}{N}$. Indeed, the total payoff of all users is maximized when everyone contributes to the public good. However, it has been proved that the Nash equilibrium in this game is defection by all users. A complete survey of PGGs and related results is available in \cite{archetti2012game}. 

Following our definition for \emph{shard-based blockchain game} $\mathbb{G}$, we show in the following theorem that  $\mathbb{G}$ is a PGG. In other words, the system fails to make any new block and remain in the same state if all processors defect initially.

\begin{theorem}\label{thm:PGGALLD}
In each epoch of a shard-based blockchain game $\mathbb{G}$ with N processors, if rewards are equally shared among all processors, then $\mathbb{G}$ reduces to a public goods game. 
\end{theorem}

\begin{proof}
Let us consider the strategy profile where all processors defect and do not pay optional cost $c^o$ after joining to the shards. We call this strategy profile $All-D$. The payoff of each processor $i$ would be then $u_i=-c^m.$ In this case, none of the processors can unilaterally change his strategy to increase its payoff. Because, the only cooperative processor cannot obtain any reward without the contribution of at least $\tau -1$  other processors in its shard, as addresses in Section \ref{sec:sysmodel}. In other words, the new payoff of each processor who deviates would be $-c^m-c^f-|x_i^j|c^v$ which is indeed smaller than $-c^m$. Hence, $All-D$ is a Nash equilibrium profile in this game and $\mathbb{G}$ is a PGG.
\end{proof}

Theorem \ref{thm:PGGALLC} further shows that we can never enforce an all-cooperation strategy ($All-C$) in the game $\mathbb{G}$, as it is not a Nash Equilibrium.

\begin{theorem}\label{thm:PGGALLC}
In each epoch of a shard-based blockchain game $\mathbb{G}$ with N processors, if rewards are equally shared among all processors, we cannot establish All-Cooperation strategy profile as a Nash equilibrium. 
\end{theorem}

\begin{proof}
We first assume that all $N$ processors have already cooperated in transaction verifications (i.e., $All-C$ strategy profile) and payed the optional cost $c^o$. We can compute the payoff of each processor $P_i$ by Equation (\ref{eq:PayoffC}). Hence, if a given processor deviates from the cooperation and play defection unilaterally, its payoff would be equal to Equation (\ref{eq:PayoffD}), which is always greater than cooperative payoffs at Equation (\ref{eq:PayoffC}). Hence, each user has incentive to deviate unilaterally and increases its payoff. Then, the $All-C$ strategy profile is never a Nash equilibrium. 
\end{proof}

Finally, Theorem \ref{thm:PGGSomeC} shows the conditions under which we can enforce an equilibrium in game $\mathbb{G}$, where some processors cooperate.

\begin{theorem}\label{thm:PGGSomeC}
Let $\mathcal{C}_j^{l_j}$ and $\mathcal{D}_j^{n-l_j}$ denote the sets of $l_j$ cooperating processors and $n-l_j$ defecting processors inside each shard $j$ with $n$ processors. If $L=\sum_{j=1}^{k}l_j$ is the total number of cooperative processors, $(\mathcal{C}^L,\mathcal{D}^{N-L})$ represents a Nash equilibrium profile in each epoch of the game $\mathbb{G}$, if and only if $l_j = \tau$ in all shards $j$, where $\mathcal{C}^L = \bigcup_{j}\mathcal{C}_j^{l_j}$ and $\mathcal{D}^{N-L} = \bigcup_{j}\mathcal{D}_j^{n-l_j}$.
\end{theorem}

\begin{proof}
If in all shards, there exist exactly $l_j = \tau$ cooperative processors, any cooperative processor cannot deviate unilaterally to increase its payoff. Because, the deviation will remove $y^j$ transaction fees from the benefits and consequently its payoff would be decreased. In the worst case, the system could even potentially fail to add a new block to the chain and all benefits would be zero. Moreover, similar to previous cases, there is no incentive to deviate for defective processors, since they must pay an extra charge for their cooperation, while this will not change the result of the consensus algorithm.
\end{proof}

The above theorems prove that if rewards are uniformly distributed among processors, a cooperative equilibria cannot be enforced in shard-based public permissionless blockchains. Hence, in the following section we define a new reward sharing approach, which promotes cooperation among processors by providing appropriate incentives.

\subsection{Fair Reward Sharing}
\label{sub:FairReward}

In this section, we extend our game model to include a fair reward sharing approach, where each processor receives a reward if and only if it has already cooperated with other processors within the shard. Let's call this new game $\mathbb{G} ^F$, in which the payoff of cooperative processors in set $\mathcal{C}^{l_j}$  is

\begin{equation}\label{eq:UC_Fair}
u_i^j(C) = \frac{BR}{kl_j} + \frac{r|y^j|}{l_j} - (c^m+c^f+|x_i^j|c^v),
\end{equation}

Recall that we assume $l_j \ge \tau$ for the consensus algorithm and each shard $j$ will submit a non-empty $y^j$ set to the blockchain. Analysis of the case where the processors cannot make a consensus on a given vector of transactions can be easily extended from our model, by assigning $BR$ and $r$ a value of zero (no benefits). As Equation (\ref{eq:UC_Fair}) shows, we first assume that the $BR$ is uniformly distributed among shards and each cooperative processor can receive a share of it. Moreover, each shard $j$ receives all fees for all transactions that it has submitted to the blockchain. Then, in each shard this reward is uniformly distributed among all cooperative processors. It is worth mentioning that $|x_i^j|$ may not always be equal to $|y^j|$. It means that a processor $P_i$ might be cooperative but finally all other processors may agree on a vector of transactions $y^j$ that is different from $x_i^j$. Thus, contrary to the standard shard-based protocols, in $\mathbb{G} ^F$ the defective processors' payoff can be calculated as 
\begin{equation}\label{eq:UD_Fair}
u_i^D = - c^m,
\end{equation}
because the defective processors will not receive any benefit. It is easy to show that the conditions of Theorem \ref{thm:PGGALLD} still hold in the new game $\mathbb{G} ^F$ and the game $\mathbb{G} ^F$ is PGG.  However, we can show that in this newly defined game $\mathbb{G}$, it is easier to enforce users to cooperate at a Nash equilibrium profile. We derive the conditions under which there exists a cooperative Nash equilibrium profile in game $\mathbb{G} ^F$, with the following theorem. 

\begin{theorem}\label{thm:PGG_Fair}
Let $\mathcal{C}_j^{l_j}$ and $\mathcal{D}_j^{n-l_j}$ denote the sets of $l_j$ cooperating processors and $n-l_j$ defecting processors inside each shard $j$ with $n$ processors, respectively. $(\mathcal{C}^L,\mathcal{D}^{N-L})$ represents a Nash equilibrium profile in each epoch of game $\mathbb{G}^F$, if the following conditions are satisfied:
\begin{enumerate}
\item In all shards $j$, $l_j \ge \tau$.
\item If for a given processor $P_i$ in shard $j$, $x_i^j = y^j$, then the number of transactions $|x_i^j|$ must be greater than $\theta_c^1 = \frac{c^f - \frac{BR}{kl_j}}{r/l_j - c^v}$  
\item If for a given processor $P_i$ in shard $j$, $x_i^j \neq y^j$, then the number of transactions $|x_i^j|$ must be smaller than $\theta_c^2 = \frac{ \frac{BR}{kl_j}+  \frac{r |y^j|}{l_j } - c^f }{c^v}$.
\end{enumerate}
\end{theorem}

\begin{proof}
The number of cooperative processors must be greater that the consensus threshold $\tau$, otherwise the cooperative processors will not receive any transaction and block reward benefits for their cooperation. Hence, they can increase their payoff by unilaterally deviating from cooperative strategy. 

We now find the largest group of cooperative processors $l_j$ in each shard, where no processor in $\mathcal{D}_j^{n-l_j}$ can join $\mathcal{C}_j^{l_j}$ to increase its payoff. Let's assume that $l_j^*$ is this largest set of processors. If processor $P_i^j$ is among the set of cooperative processors, then it will not unilaterally deviate if its payoff (calculated by Equation (\ref{eq:UC_Fair})) is greater than $-c^m$. Two possible cases could happen in this case. First, $P_i^j$ could be among processors who have the same vector of transactions as the output of the shard, i.e., $x_i^j = y^j$. In this case, $P_i^j$ will not deviate from cooperation if:

\begin{equation}\label{eq:Cooper1}
 \frac{BR}{kl_j} + \frac{r|x_i^j|}{l_j} - (c^m+c^f+|x_i^j|c^v) \ge -c^m,
\end{equation}

which shows that $x_i^j \ge \theta_c^1,$ where  $$\theta_c^1= \frac{c^f - \frac{BR}{kl_j}}{r/l_j - c^v}. $$

In the second case, processor $P_i^j$ have cooperated with others in $\mathcal{C}_j^{l_j}$, but its vector of transactions is different from the output of the shard, i.e., $x_i^j \neq y^j$.  Hence, the following condition must be satisfy if this user wants to remain in the cooperative set.

\begin{equation}\label{eq:Cooper2}
 \frac{BR}{kl_j} + \frac{r|y^j|}{l_j} - (c^m+c^f+|x_i^j|c^v) \ge -c^m, 
\end{equation}

which shows that $x_i^j < \theta_c^2,$ where  $$\theta_c^2= \frac{ \frac{BR}{kl_j}+  \frac{r |y^j|}{l_j } - c^f }{c^v}. $$ If $l_j^*$ represents the largest set of cooperative processors in each shard, then $(\mathcal{C}^L,\mathcal{D}^{N-L})$ would be the unique cooperative Nash equilibrium of the game $\mathbb{G} ^F$.
Please note that this NE is a unique cooperative equilibrium of the game, as we have already found the largest set of cooperative processors in all shards.
\end{proof}

Note that  by increasing the optional costs of computation (whether $c^f$ is in the numerator or $c^v$ in denominator of $\theta_C$) and for any given number of transactions $|x_i^j|$, processors will be tempted to be more defective as the threshold $\theta_c^1$ will be increased. This is in line with our intuition that processors are not cooperative if the cost of cooperation is high. On the other hand, the calculated threshold shows that by increasing the number of processors $N$ and consequently the number of shards $k$, the processors would be more defective. This is representing the case where the processors will not cooperate in the hope that other processors will participate in the transaction verifications and other optional tasks in the defined protocol. Moreover, as the reward is smaller, the processor must obtain more benefits from the transaction fees to have positive payoff.  In other words, cooperative processors have less incentives to cooperate, because the number of participants is more and they will receive smaller reward. 

Recall that the game $\mathbb{G}^F$ is still a social dilemma game, but with the new reward distribution approach we can provide enough incentives to enable processor cooperation. Our results in this section showed that a shard-based permissionless blockchain protocol could be potentially a PGG and processors could remain in $All-D$ equilibrium without any reward. We also showed that the cooperation can be enforced under some conditions where the number of transactions are large enough for the  processors. Next we apply the results from Theorem \ref{thm:PGG_Fair} to design an \emph{incentive-compatible} sharding protocol for public permissionless blockchain.
\begin{algorithm}[ht]
	\footnotesize
	\caption{Incentive-Compatible Protocol}\label{Algo-BC}
	\begin{algorithmic}[5]
		\Procedure{Initialization and Committee Creation}{}
		\State { $ID, Shard \gets ComputeID(epochRandomness, IP, PK)$}
		\State { $x_i \gets ShardTransactions(Shard)$}
		\EndProcedure
		
		\Procedure{Cooperaive/Defective Node Selection}{}
		\State $P_i$ sends $H(x_i^j)$ to $Coordinator$
		\If{$Coordinator$}
		\State Receive $H(x_i^j)$s
		\State $l_j \gets$  Maximum number of processors with \\ \quad \quad \quad \quad \quad \quad common  transactions
		
		\If{$l_j < \tau$}
		\State \Return $All-D$
		\Else
		\State Prepare the list of $l_j$ processors $\mathcal{C}_j^{l_j}$ 
		\State Calculate $\theta_c^1$ and $\theta_c^2$ from \textit{Theorem 4}
		\State \Return $\theta_c^1$, $\theta_c^2$, and $\mathcal{C}_j^{l_j}$
		\EndIf
		\EndIf
		\EndProcedure
		
		\Procedure{Shard Participation (Consensus)}{}
		\If{$P_i \in  \mathcal{C}_j^{l_j}$ and $|x_i^j| \le \theta_c^1$}
		\State \Return Defect
		\ElsIf{$P_i \notin  \mathcal{C}_j^{l_j}$ and $|x_i^j| \ge \theta_c^2$}
		\State \Return Defect
		\EndIf
		\State Verify transactions and create a set of verified transactions $y^j$ by all remaining cooperative processors 
		\State Consensus on verified transactions
		\State Sign BFT agreement result
		\State \Return Signature, Agreed block's header
		\EndProcedure
		
		\Procedure{Verification, Reward, and Punishment}{}
		\State Verify whether $P_i \in C^L$ have cooperated in each shard
		\State Distribute rewards among cooperative $P_i$ according to \\ \quad \quad Equation~(\ref{eq:UC_Fair})
		\EndProcedure
	\end{algorithmic}
\end{algorithm}

\section{Incentive-Compatible Reward Sharing}
\label{sec:MD}

As discussed earlier, in any shard-based protocol, processors may not have enough incentives to cooperate and verify transactions, which leads them to a social-dilemma. In other words, the decision of each processor (to cooperate or defect) exclusively depends on the number of received transactions compared to a fixed threshold. 
Our game-theoretic evaluation allows us to design a more sophisticated protocol - the \emph{incentive-compatible} reward sharing - that extends current shard-based protocols by considering optimal strategies of processors and enforcing cooperation in these protocols. The \emph{incentive-compatible} reward sharing protocol is based on our results for fair reward distributions presented in Section~\ref{sec:results}.

Algorithm \ref{Algo-BC} outlines the main steps of our proposed protocol. Comparing to the standard shard-based protocols, the main difference of our proposed \emph{incentive-compatible} protocol is that we first announce to processors whether the cooperation would be in their interests. The protocol proceeds as follows. The processors first try to solve the PoW puzzle and obtain an ID to participate in a committee. After committee formation and assignment, each processor $P_i$ in shard $j$ receives a list of $x_i^j$ transactions to verify.

At this stage, all processors calculate the $H(x_i^j)$ and submit it to a coordinator, where $H$ is a predefined hash function. Note that the coordinator could be potentially one of the processors that has been selected randomly in each shard. This coordinator could be a centralized trusted third party as well, as it does not receive any sensitive information. The coordinator then finds the maximum subset of processors with similar $H(x_i^j)$. This will estimate $l_j$ and $\mathcal{C}_j^{l_j}$, which will be then used to calculate the $\theta_c^1$ and $\theta_c^2$.

As described in Algorithm \ref{Algo-BC}, the \emph{incentive-compatible} protocol assists processors in selecting the optimal strategy in a given epoch. We assume that the set of cooperative processors $\mathcal{C}_j^{l_j}$ and the set of defective processors $\mathcal{D}_j^{n-l_j}$ are publicly announced in each shard. 
In each epoch, the protocol defines publicly which rational processor must cooperate or defect, by considering the  number of received transactions by all processors and based on the results in Theorem \ref{thm:PGG_Fair}. At the end of committee participation phase, it is easy to verify if a given processor $P_i$ has already followed the recommendations by the \emph{incentive-compatible} protocol or they have deviated from the defined strategies. In case a processor in $\mathcal{C}_j^{l_j}$ has not cooperated in this phase, the \emph{incentive-compatible} protocol will not give this processor any reward at the end of this epoch. This is also a novel punishment approach that has been added to our proposed protocol compared to previous ones.

\begin{figure*}[t]
\centering
\begin{subfigure}{0.24\linewidth}
\centering
\includegraphics[width=\textwidth]{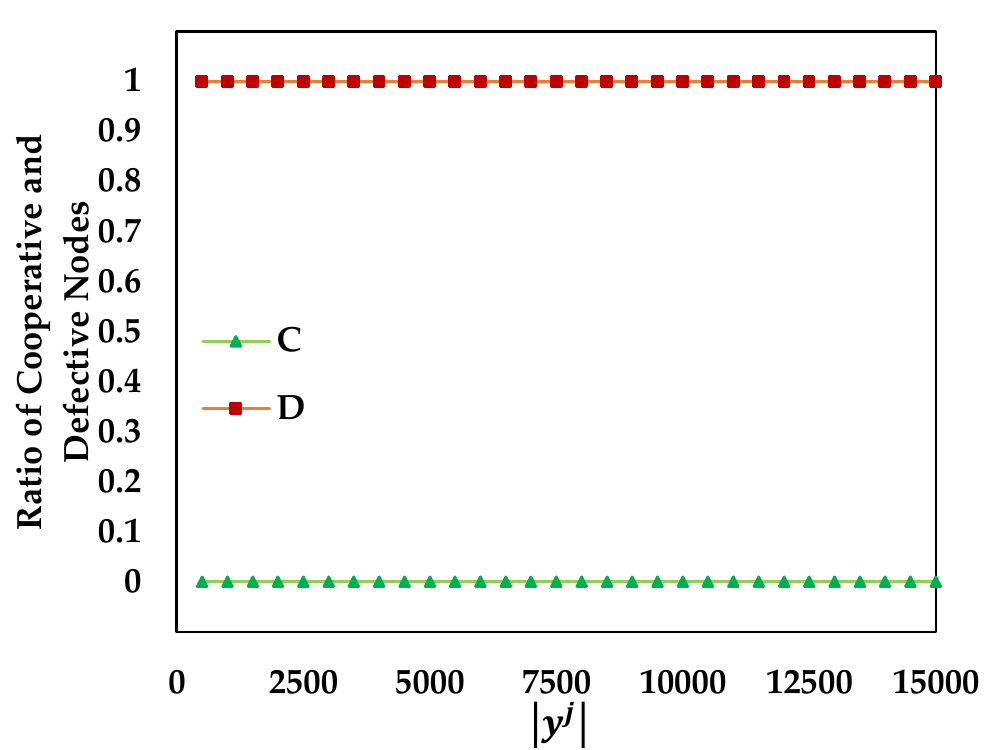}
\caption{Uniform}
\label{uniform-y^j}
\end{subfigure}
\begin{subfigure}{0.24\linewidth}
\centering
\includegraphics[width=\textwidth]{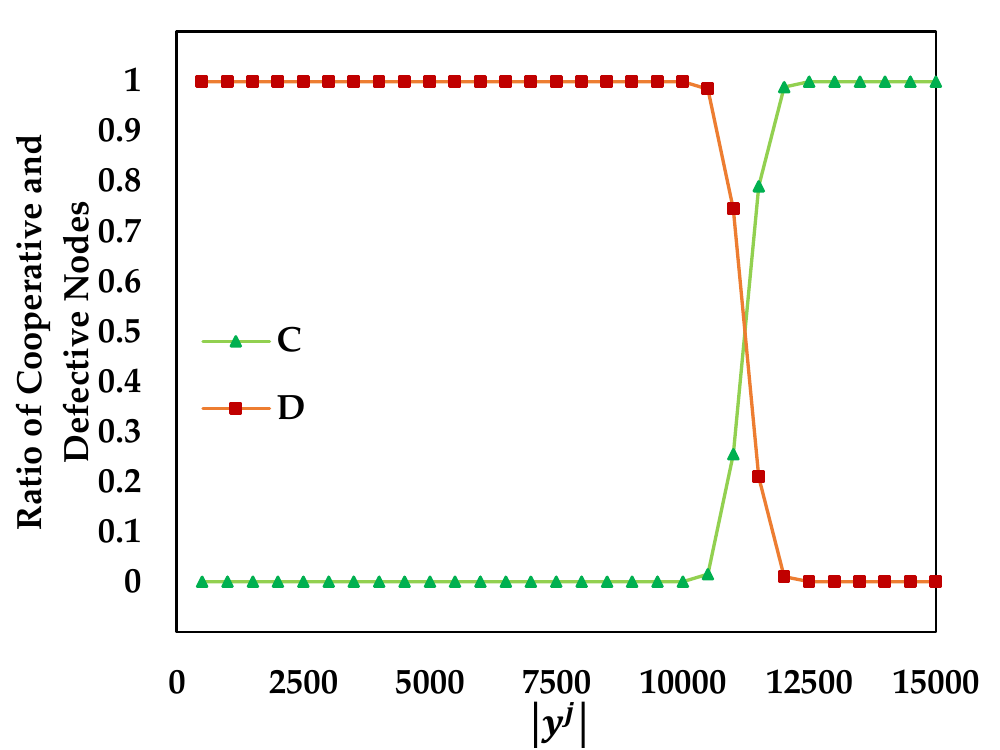}
\caption{Fair}
\label{fair-y^j}
\end{subfigure}
\begin{subfigure}{0.24\linewidth}
\centering
\includegraphics[width=\textwidth]{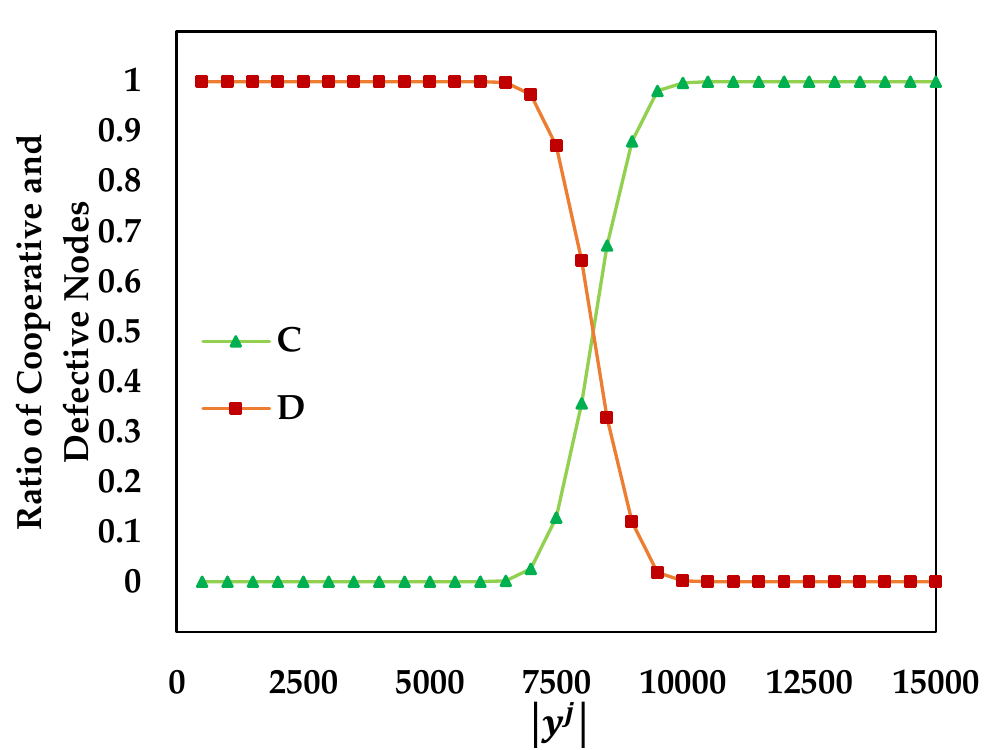}
\caption{Incentive-Compatible}
\label{incentive-compatible-y^j}
\end{subfigure}
\begin{subfigure}{0.24\linewidth}
\centering
\includegraphics[width=\textwidth]{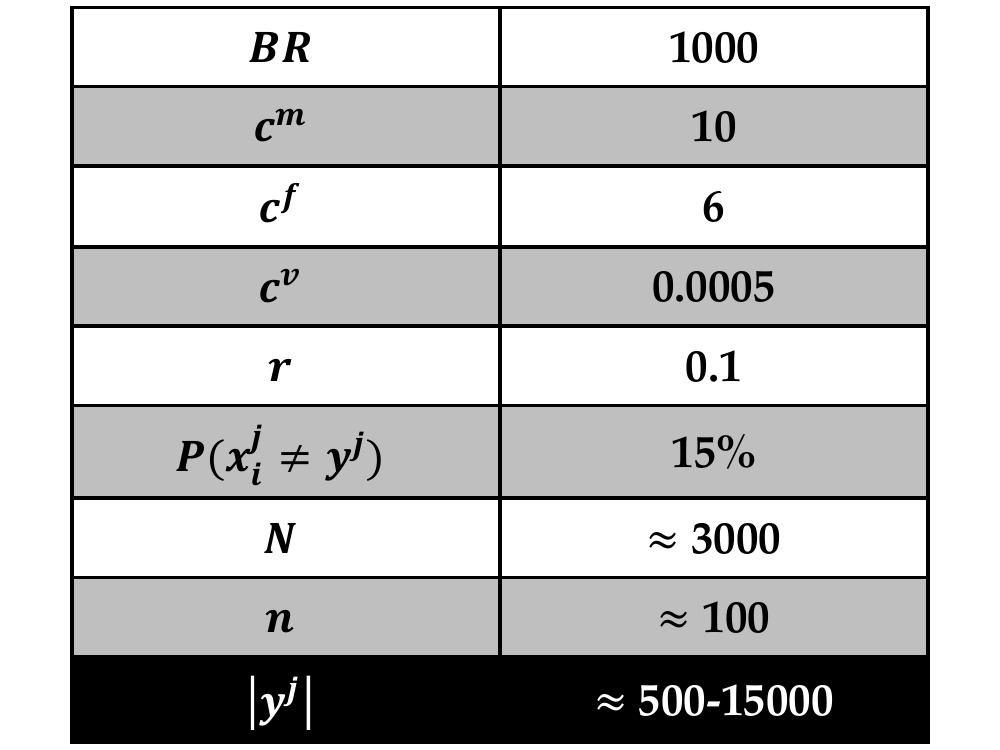}
\caption{Simulation Parameters}
\label{parameters-y^j}
\end{subfigure}
\caption{Ratio of \emph{cooperative} and \emph{defective} processors for different sizes of $y^j$.}
\label{results-y^j}
\end{figure*}

\begin{figure*}[t]
\centering
\begin{subfigure}{0.24\linewidth}
\centering
\includegraphics[width=\textwidth]{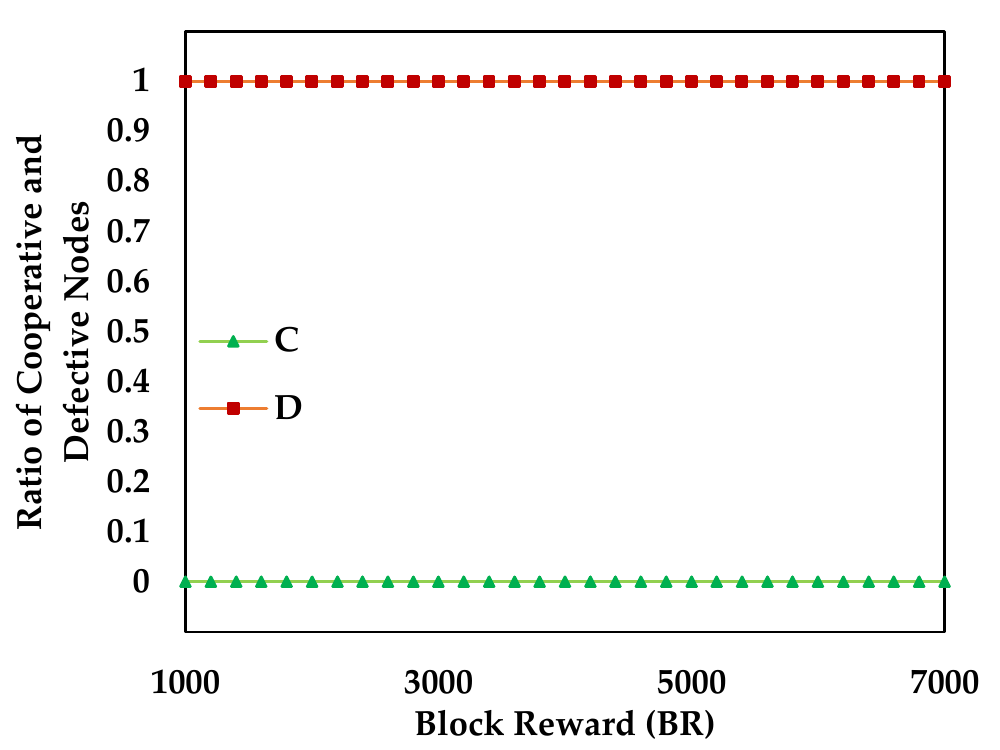}
\caption{Uniform}
\label{uniform-BR}
\end{subfigure}
\begin{subfigure}{0.24\linewidth}
\centering
\includegraphics[width=\textwidth]{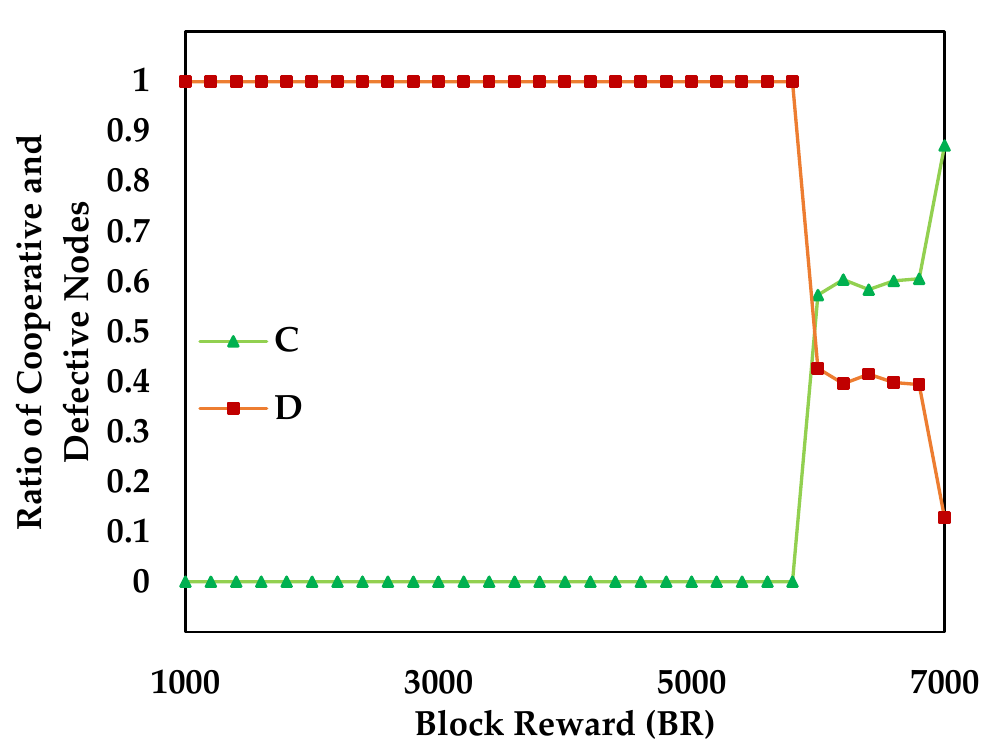}
\caption{Fair}
\label{fair-BR}
\end{subfigure}
\begin{subfigure}{0.24\linewidth}
\centering
\includegraphics[width=\textwidth]{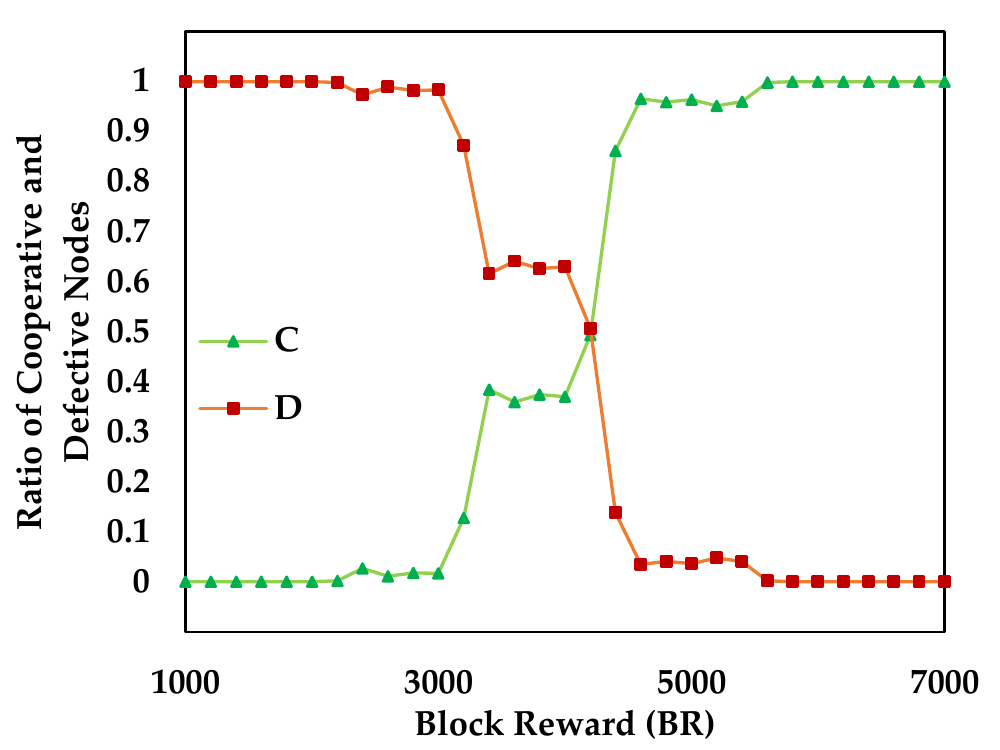}
\caption{Incentive-Compatible}
\label{incentive-compatible-BR}
\end{subfigure}
\begin{subfigure}{0.24\linewidth}
\centering
\includegraphics[width=\textwidth]{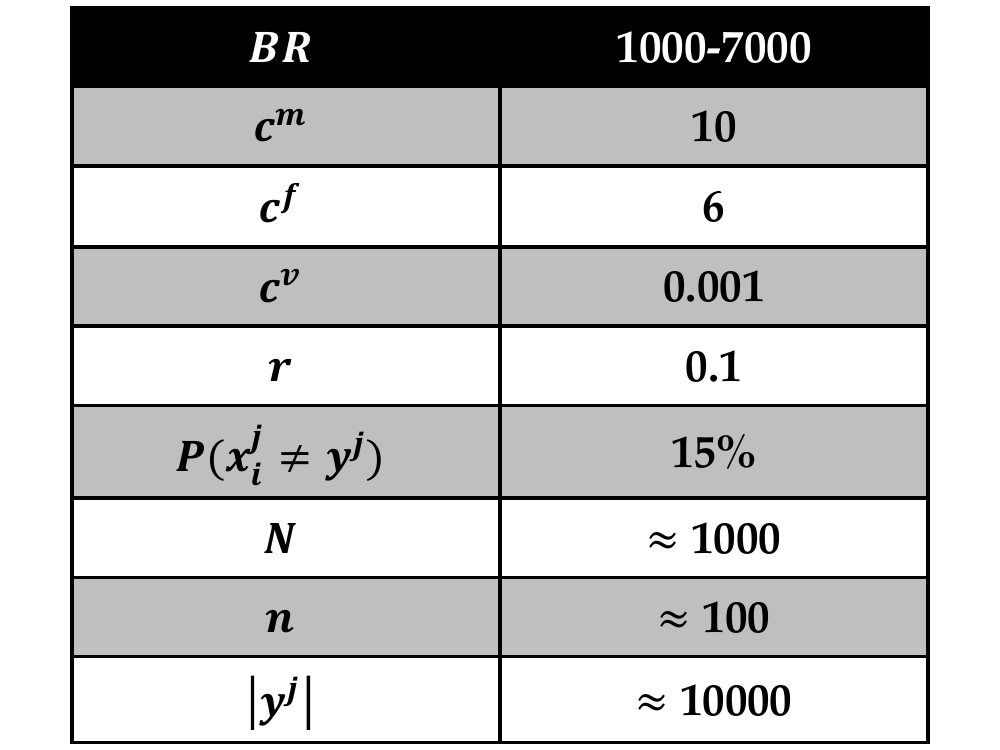}
\caption{Simulation Parameters}
\label{parameters-BR}
\end{subfigure}
\caption{Ratio of \emph{cooperative} and \emph{defective} processors for different values of $BR$.}
\label{results-BR}
\end{figure*}

\section{Numerical Analysis}
\label{sec:Sim}

We conduct a comprehensive set of numerical simulations, in order to validate how our proposed \emph{incentive-compatible} protocol compares with uniform and fair reward sharing protocols in shard-based blockchains. We first detail the experimental setup used to simulate a basic shard-based blockchain in Section \ref{expsetup}. Variants of the simulation were used to analyze multiple parameters that may affect the strategy of individual processors, and thereby its effect on the successful operation of the blockchain network.

\subsection{Experimental Setup}
\label{expsetup}
We simulate a shard-based public permissionless blockchain with approximately $N$ $(\pm1\%)$ processors that are selfishly following a protocol to reach consensus in each shard, then combine all shards to add the next block, and finally collect their reward at the end of each epoch. We assume committees of size $100$ $(\pm1\%)$, and the required number of processors in each shard for consensus is $\tau\approx51$. Also, the number of committees (and shards) grow linearly \emph{w.r.t.} the number of processors in the network ($k\approx\frac{N}{100}$). Each processor in the network is assumed to receive $|x_i^j|\approx|y^j|$ $(\pm1\%)$ transactions corresponding to the shard it belongs to. As imperfect views of the network is common occurrence in real-world networks, we also assume that the number of processors with $x_i^j\neq{y^j}$ is approximately 15\%. We present mean results of 100 iterations for each combination of parameters (in Figures \ref{results-y^j}-\ref{results-uN}), i.e., every point in the graphs was obtained after averaging the results of 100 independent epochs with that particular set of parameters.

\subsection{Number of Transactions}
We first analyze the effect of varying the average number of transactions $|x_i^j|$ between $500$ and $15000$. The corresponding ratios of cooperative and defective processors is plotted in Figure \ref{results-y^j}. As intuitive, the uniform reward sharing results in all defect (Figure \ref{uniform-y^j}), and thus no block is ever added to the blockchain. In case of fair and incentive-compatible reward sharing protocols (Figure \ref{fair-y^j} and \ref{incentive-compatible-y^j}, respectively) we observe that processors opt for all defect strategy when the number of transactions is low, but eventually change their strategy to cooperate as the number of transactions gets high enough to make a profit. More importantly, the proposed incentive-compatible reward sharing protocol achieves a majority of cooperative processors for lesser number of transaction than in the case of fair sharing, which is favorable.

\subsection{Block Reward}
We next analyze the effect of varying the block reward $BR$ between $1000$ and $7000$, and the corresponding ratios of cooperative and defective processors is plotted in Figure \ref{results-BR}. As before, the uniform reward sharing results in all defect (Figure \ref{uniform-BR}), regardless of the value of the block reward. In case of fair and incentive-compatible reward sharing protocols (Figure \ref{fair-BR} and \ref{incentive-compatible-BR}, respectively) we observe that processors opt for all defect strategy when the block reward is low, but eventually change their strategy to cooperate as the block reward gets high enough to make a profit. Again, the proposed incentive-compatible reward sharing protocol achieves a majority of cooperative processors for lesser valued block reward than in the case of fair sharing, which is favorable.

\subsection{Size of the Network}
\begin{figure*}[t]
\centering
\begin{subfigure}{0.24\linewidth}
\centering
\includegraphics[width=\textwidth]{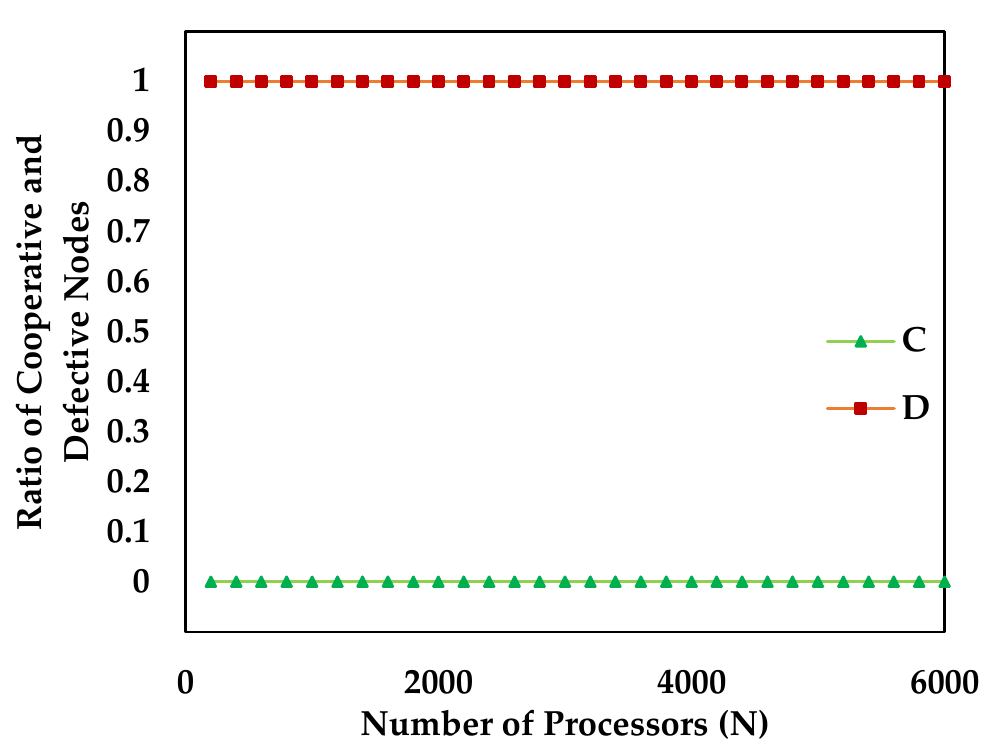}
\caption{Uniform}
\label{uniform-N}
\end{subfigure}
\begin{subfigure}{0.24\linewidth}
\centering
\includegraphics[width=\textwidth]{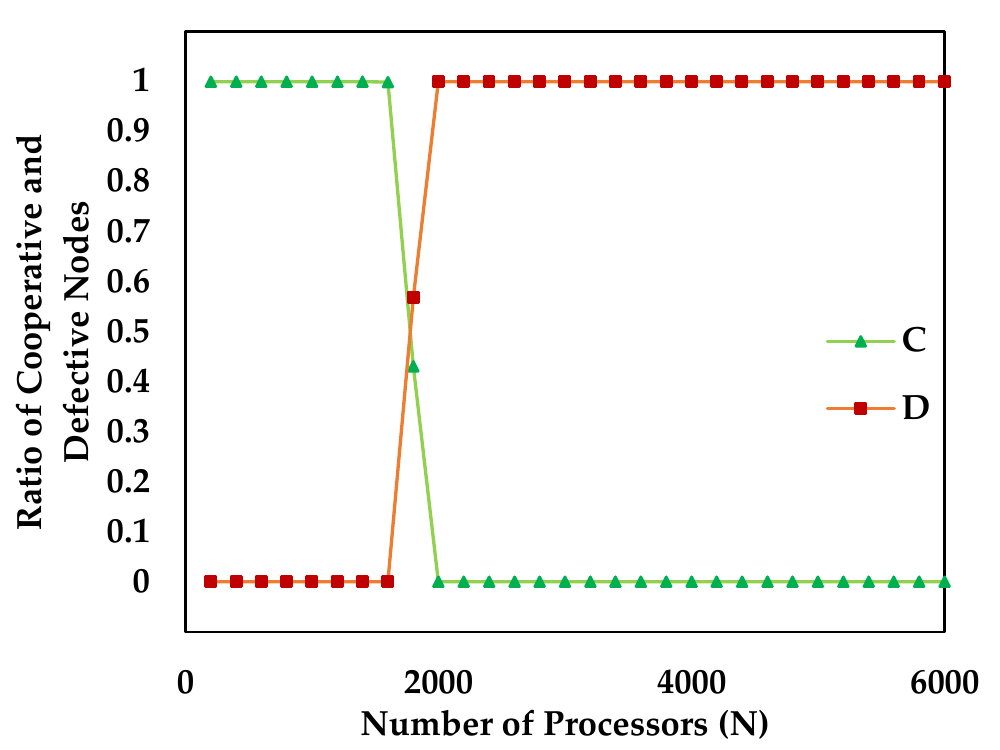}
\caption{Fair}
\label{fair-N}
\end{subfigure}
\begin{subfigure}{0.24\linewidth}
\centering
\includegraphics[width=\textwidth]{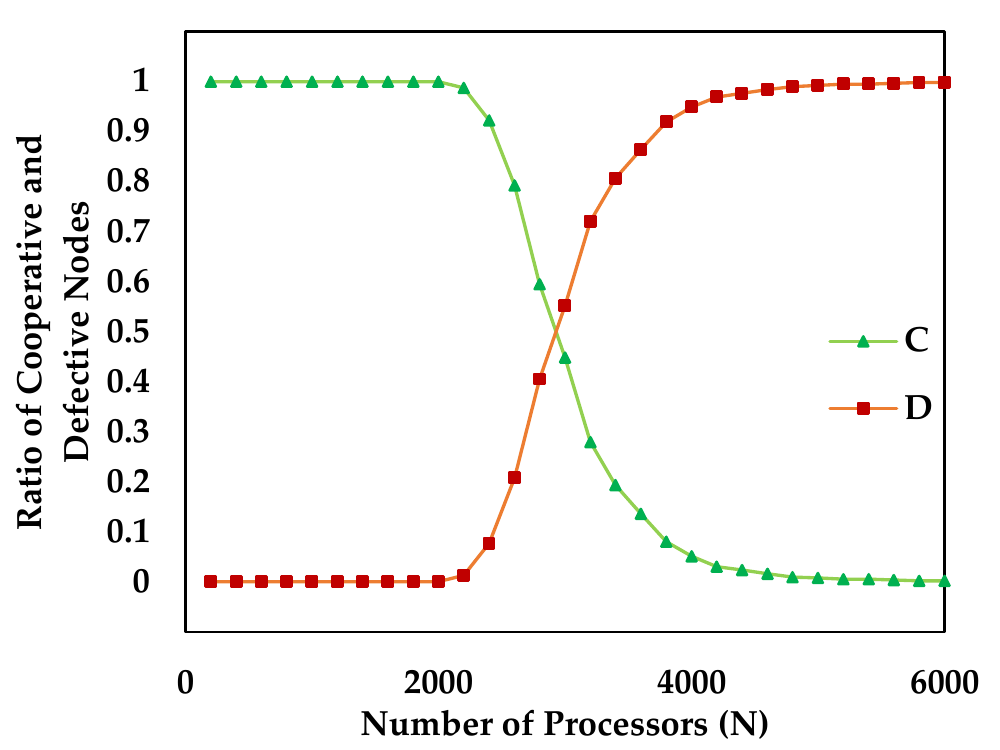}
\caption{Incentive-Compatible}
\label{incentive-compatible-N}
\end{subfigure}
\begin{subfigure}{0.24\linewidth}
\centering
\includegraphics[width=\textwidth]{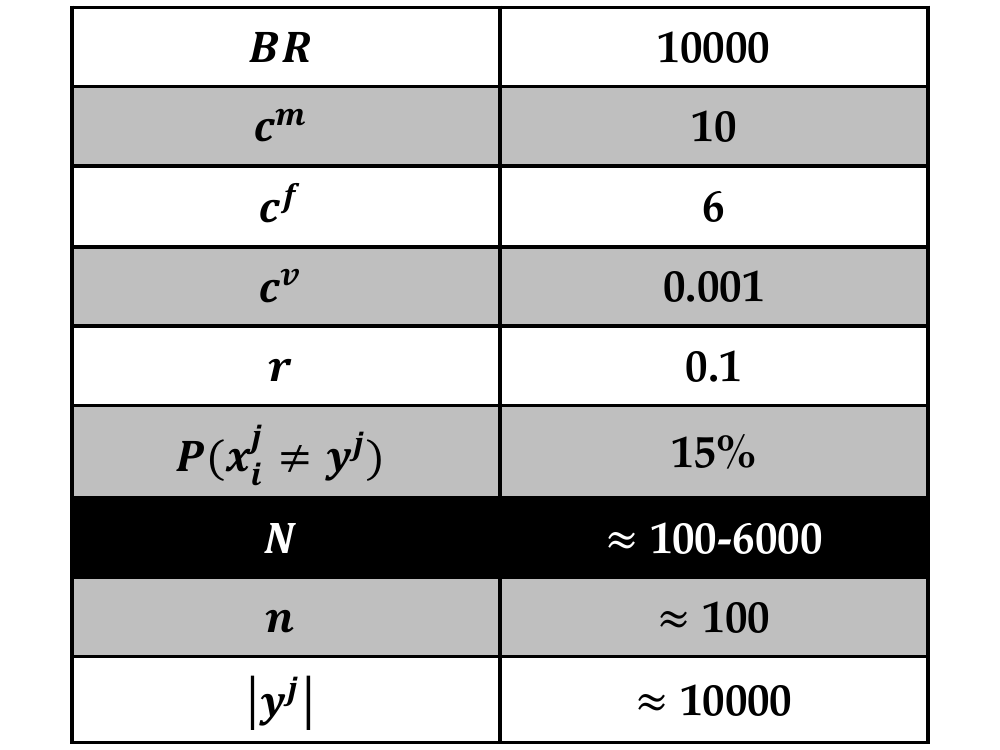}
\caption{Simulation Parameters}
\label{parameters-N}
\end{subfigure}
\caption{Ratio of \emph{cooperative} and \emph{defective} processors for different values of $N$.}
\label{results-N}
\end{figure*}

The number of processors in the network in a given epoch can vastly impact the strategy for individual processors, because if a small reward is shared between a large number of cooperative processors, it may not cover other costs associated with participation (such as $c^f$). We observe this intuition in effect in Figure \ref{results-N}, where $N$ is varied between  $100$ and $6000$. Both the proposed incentive-compatible and fair reward sharing protocols lose majority of cooperative processors when $N$ is increased significantly. However, the proposed incentive-compatible reward sharing protocol retains a majority of cooperative processors for greater number of processors than in the case of fair sharing, which is desirable. As before, the uniform reward sharing results in all defect (Figure \ref{uniform-N}), regardless of the number of processors.

\begin{figure*}[t]
\centering
\begin{subfigure}{0.24\linewidth}
\centering
\includegraphics[width=\textwidth]{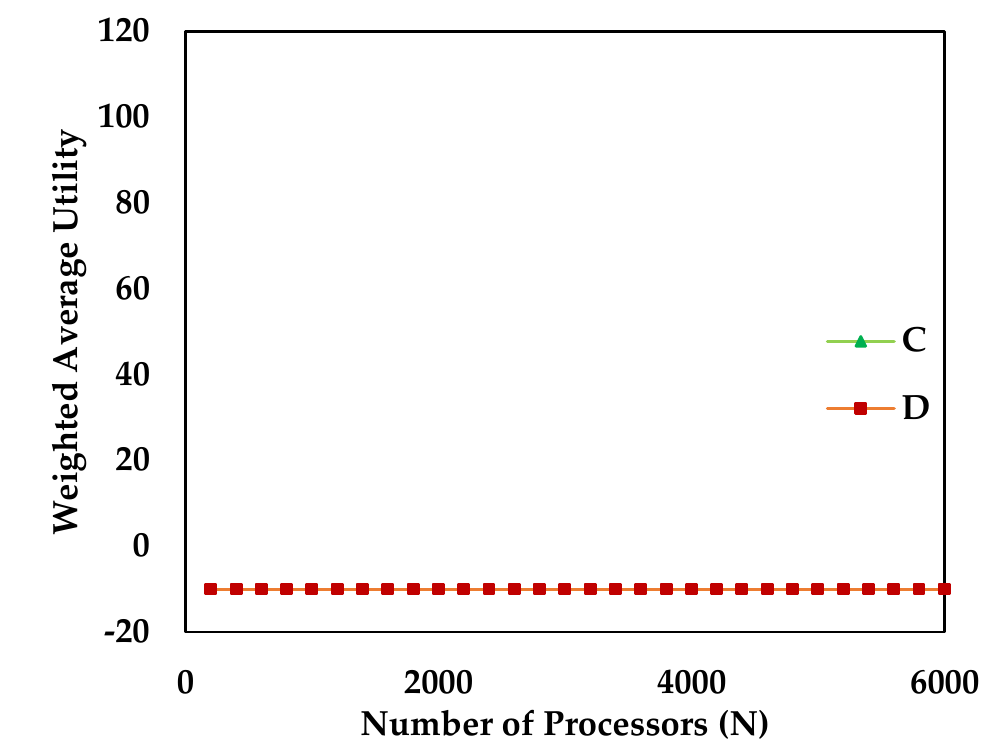}
\caption{Uniform}
\label{uniform-uN}
\end{subfigure}
\begin{subfigure}{0.24\linewidth}
\centering
\includegraphics[width=\textwidth]{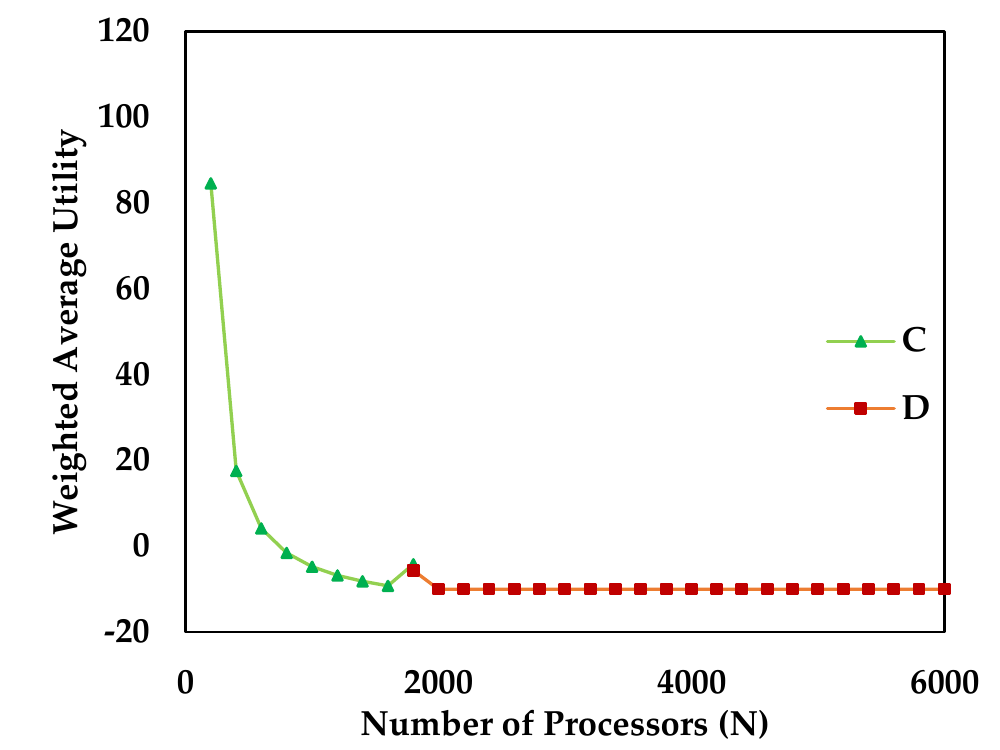}
\caption{Fair}
\label{fair-uN}
\end{subfigure}
\begin{subfigure}{0.24\linewidth}
\centering
\includegraphics[width=\textwidth]{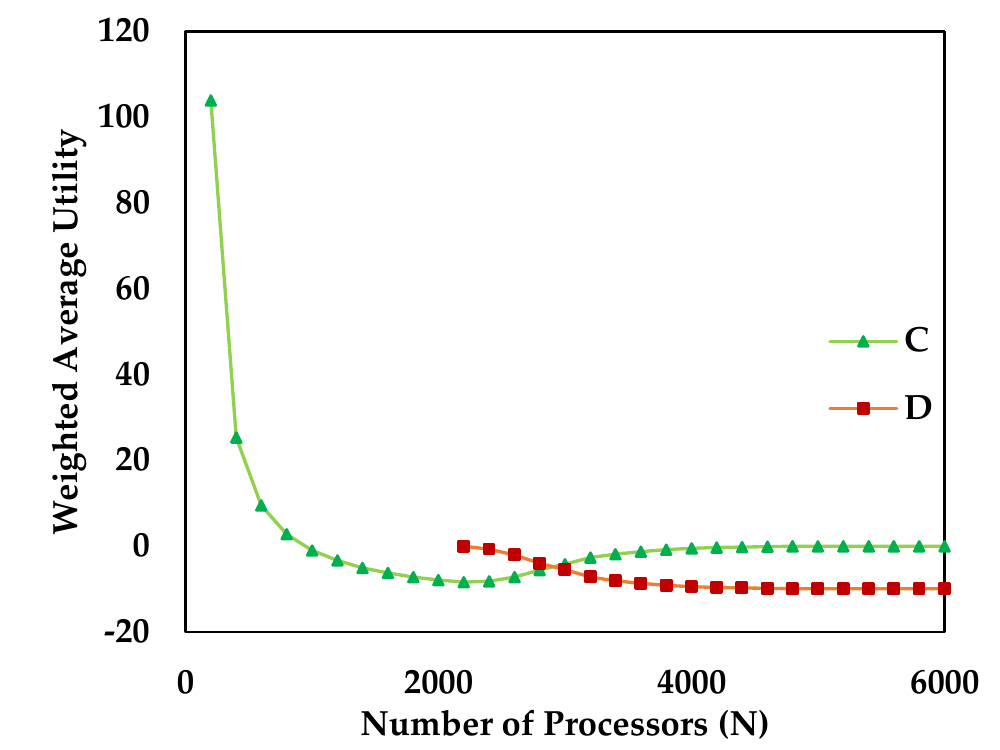}
\caption{Incentive-Compatible}
\label{incentive-compatible-uN}
\end{subfigure}
\begin{subfigure}{0.24\linewidth}
\centering
\includegraphics[width=\textwidth]{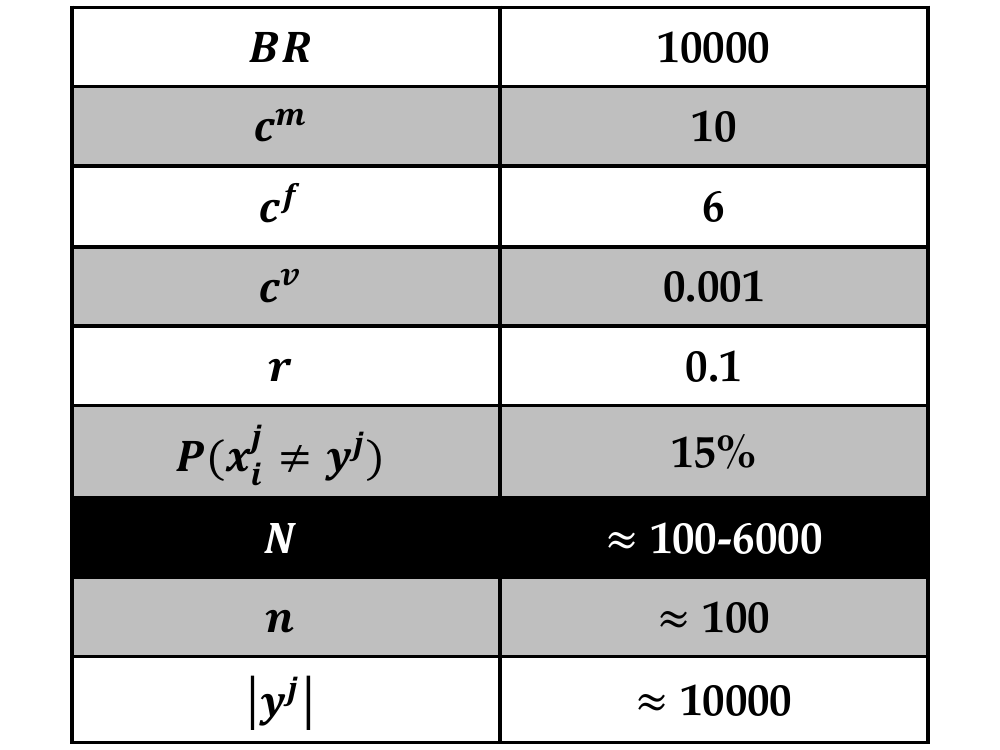}
\caption{Simulation Parameters}
\label{parameters-uN}
\end{subfigure}
\caption{Weighted average utility of \emph{cooperative} and \emph{defective} processors for different values of $N$.}
\label{results-uN}
\end{figure*}

In order to better understand why cooperative processors flip to being defective, we also plotted the corresponding weighted utility of processors in Figure \ref{results-uN}. In case of both fair and incentive-compatible protocols, the average utility drops significantly with increasing number of processors. The average utility gradually converges at about $-c^m$ (which is $-10$ in our simulation). As utility by cooperation drops below $-c^m$, processors flip to being defective and incur only $-c^m$. Also, in uniform reward sharing we see a constant $-c^m$ utility for all (defective) processors.

\subsection{Discussion and Limitations}
Our goal in this work was to design practical incentive mechanisms for eliciting cooperation in shard-based blockchains. The above analytical and empirical results shows how our proposed reward sharing mechanism promotes cooperation in shard-based blockchains, and thwarts free-riding processors. Nonetheless, there exists certain limitations in the proposed mechanism, discussed below, some of which we plan to address in the future.

\noindent
\textbf{Inter-Shard Communication.}
Due to the lack of communication between committees, cooperative processors in a shard where consensus is reached, can suffer when another committee fails to reach consensus (because no block is added to the blockchain if one or more shards fail). This can be resolved if an inter-shard communication is established in Algorithm \ref{Algo-BC}, wherein coordinators can exchange consensus status and inform potentially cooperative processors about the state of consensus in other shards as well. We plan to include inter-shard communication in our future work, and analyze how the game changes due to it.

\noindent
\textbf{Inclusion of Malicious Processors.}
In this work we consider only honest but greedy (or selfish) processors (each trying to maximize its utility) who would follow the instructions of a coordinator. However in real-world, malicious processor(s) may also exist whose sole objective may be to disrupt the blockchain network. Such malicious processors may misbehave at various stages of the protocol, such as reporting false $H(x_i^j)$ or not following coordinator's instruction to cooperate (or defect). As part of our future work, we plan to include malicious processors in the game and re-analyze the game.

\noindent
\textbf{Parametric Values.}
The parametric values chosen for our numerical analysis was primarily to showcase the trends observable across the three different reward sharing mechanisms. They may or may not be reflective of values in a real shard-based blockchain network, but we did our best to establish the inequalities between parameters as completely as possible.


\section{Related Work}
\label{sec:related}
In this section, we briefly outline the efforts in the literature towards improving the scalability and transaction rate of consensus protocols in public permissionless blockchains. For an exhaustive survey of blockchain consensus protocols in the literature, readers are referred to \cite{Bano:2017}. The original \emph{Nakamoto consensus} protocol \cite{bitcoin:2009} of \emph{Bitcoin} which employed a leader selection using PoW puzzles (to commit the next block) suffered from poor scalability and transaction throughput. \emph{Bitcoin-NG} \cite{eyal2016bitcoin} attempted to improve Bitcoin's performance by employing \emph{microblocks}. In Bitcoin-NG, similar to Bitcoin, a leader is selected using PoW in each epoch. However, unlike Bitcoin, the leader can continue to append microblocks (containing transactions) to the blockchain for the duration of its epoch, until a new leader is elected.

As leader or single node based (implicit) consensus algorithms such as Nakamoto consensus and Bitcoin-NG still suffer from poor performance, fault-tolerance and consistency issues, the community's focus shifted on designing blockchain consensus protocols using a \emph{committee} of nodes, rather than a single node (or leader). While committee-based consensus algorithms were introduced more than two decades ago \cite{bracha1987log}, much recently Decker et al. \cite{decker2016bitcoin} proposed one of the first committee-based consensus protocols for public blockchains, named \emph{PeerCensus}. However, PeerCensus did not clarify how committee formation is done and how an honest majority can be ensured within the committee. Follow up works \cite{AbrahamMNRS16,pass2017hybrid,kogias2016enhancing,gilad2017algorand} in similar direction improved the practicality of such single committee-based consensus protocols by proposing different strategies on how unbiased committees can be formed. 

Although single committee consensus algorithms provide significantly improved performance compared to single node or leader-based consensus algorithms, one major limitation of such techniques is that they do not scale well. Moreover, increasing committee size in such techniques comes at the expense of a decreased throughput. This motivated the design of blockchain consensus protocols that employ \emph{multiple committees}. The main idea in these protocols is to split the transactions among multiple committees (or shards), which then process these shards or set of transactions in parallel. This also improves the overall scalability of the system. \emph{RSCoin} \cite{DanezisM:RSCoin} was proposed as a shard-based blockchain technique for centrally-banked cryptocurrencies, while \emph{Elastico} \cite{luu2016secure} was the first shard-based consensus protocol for public blockchains. \emph{Omniledger} \cite{kokoris2017omniledger} and \emph{Rapidchain} \cite{zamanirapidchain} are some of the recently proposed shard-based public blockchain protocols that attempt to address the scalability and security issues of Elastico. Despite the recent interest in shard-based protocols for improving transaction throughput and scalability in public blockchains, there have been no prior efforts in the literature, until this one, that study the rational behavior of processors or miners in such a multiple committee approach. 



\section{Conclusions}
\label{sec:conclusions}

In this paper, we comprehensively studied the problem of selfishness in shard-based permissionless blockchains. We first introduced a system model to capture the main operational parameters in current shard-based blockchain protocols. Next, we evaluated the strategic behavior of processors in such protocols by employing concepts from game theory. Specifically, we modeled shard-based blockchain protocols as $n$-player non-cooperative games using different reward sharing scenarios and obtain the Nash equilibria (NE) strategy profile for each scenario. Based on our analytical results under different reward sharing scenarios, we designed an incentive mechanism for shard-based blockchain protocols which would enforce cooperation among processors by guaranteeing optimal incentive distribution. Our numerical analysis also validated that the proposed reward sharing mechanism outperforms uniform reward sharing and provides more incentive for cooperation when the block reward or number of transactions is small. This work is the first step towards a deeper understanding of the effect of non-cooperative behavior in shard-based blockchains.

\balance 
\bibliography{realistico-bib} 
\bibliographystyle{ieeetr}

\end{document}